# SIMULATION OF TOPCON/PERC HYBRID BOTTOM STRUCTURE FOR PEROVSKITE/SILICON TANDEM SOLAR CELLS USING QUOKKA3

Eni Muka[1], Raşit Turan[1,2,3], Hisham Nasser[1]

[1]ODTÜ-GÜNAM, Middle East Technical University, Ankara, 06800, Türkiye
[2]Micro and Nanotechnology Department, Middle East Technical University, Ankara, 06800, Türkiye
[3]Department of Physics, Middle East Technical University, Ankara, 06800, Türkiye

ABSTRACT: This work emphasizes the potential of perovskite/silicon tandem solar cells for increased power conversion efficiencies. By employing crystalline silicon (c-Si) as the bottom cell, particularly with p-type PERC technology, there are cost-effective and advantageous physical properties. However, traditional phosphorus-doped emitters in PERC Si bottom cells are hindered by high surface recombination, which limits their performance. This research introduces a novel hybrid PERC/TOPCon structure that integrates a phosphorus-doped poly-Si (n+ TOPCon) layer as the front emitter to address these challenges. Numerical simulations using Quokka3 confirmed the feasibility of the design, focusing on optimizing the rear side metallization to enhance implied open-circuit voltage ($V_{oc}$) and fill factor (FF). A two-step process systematically varied local contact openings to examine their impact on performance metrics. Results highlighted optimal rear metallization parameters, achieving optimal metal fractions approximately 2%. This innovative approach demonstrates the effectiveness of combining TOPCon and PERC technologies for bottom cells in tandem structures, providing valuable insights into their development and optimization. The study underscores the potential of the hybrid PERC/TOPCon structure in enhancing the functionality and efficiency of perovskite/silicon tandem solar cells.

## 1 INTRODUCTION

The progression toward enhanced solar efficiencies through tandem cells is remarkable, particularly with the rapid advancements in the last decade [1,2]. The choice of crystalline silicon (c-Si) for perovskite/silicon tandem solar cells is promising due to its advantageous physical properties and cost-effectiveness, and optimizing c-Si as the bottom cell is crucial for improving the overall functionality and efficiency of such devices [3]. Given its widespread adoption and market presence, the utilization of p-type PERC technology as the foundational concept for the bottom cell in perovskite/silicon tandem devices is reasonable. However, conventional phosphorus-doped emitters in PERC Si bottom cells can lead to high surface recombination, presenting a challenge [4]. To address this challenge, a novel solution integrates a PERC structure with a phosphorus-doped poly-Si (n+ TOPCon) layer as the front emitter, resulting in a hybrid PERC/TOPCon structure. This innovative approach shows promise in overcoming the limitations of conventional PERC-based bottom cells in perovskite/silicon tandem cells [5]. This work focuses at developing a viable PERC- based bottom cell structure with front TOPCon for perovskite/silicon tandem solar cells are presented. Using numerical simulations conducted by the Quokka3 software, the aim is to validate the feasibility and performance potential of this design. Our investigation emphasizes optimizing the rear side metallization design of the PERC/TOPCon structure, intending to enhance key performance metrics such as open-circuit voltage ($V_{oc}$) and fill factor (FF). By systematically varying the local contact opening (LCO) and analyzing its impact on device performance, the development of a comprehensive roadmap for achieving optimal results has been put into focus. Overall, this study aims to contribute valuable insights into developing and optimizing bottom-cell structures for perovskite/silicon tandem solar cells. This study shows that combining TOPCon and PERC technologies is effective for the bottom cell in a tandem structure.

## 2 METHODS

Simulations were conducted on a 2.2 x 2.2 cm² hybrid PERC/TOPCon structure, with the specific layer compositions and parameters outlined in Figure 1 and Table I, all of which were measured experimentally in our lab.

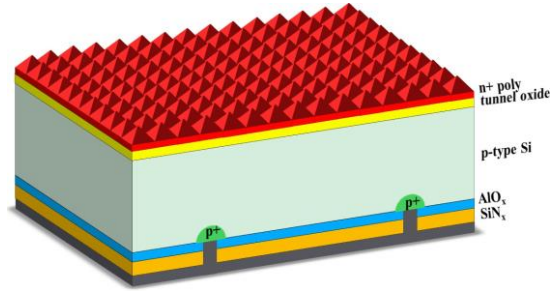

**Figure 1:** Simulated Hybrid TOPCon/PERC Structure

**Table I:** Simulation Parameters for Quokka 3

| Type of Parameters | | Value |
|---|---|---|
| Optical Parameters | Front Side Transmission | Obtained by OPAL 2 |
| Bulk | Thickness<br>Resistivity<br>Lifetime | 150 μm<br>1 Ωcm<br>1 ms |
| Front Side Emitter | $J_{TOPCon}$<br>$R_{sheet}$ | 3 fA/cm2<br>60 Ω/□ |
| Local BSF | $R_{sheet}$<br>$J_{0,metal}$<br>$J_{0,passivated}$ | 40 Ω/□<br>1126.24 fA/cm2<br>5 fA/cm2 |

While the front side has no metal, the rear side has local metal contacts, forming the back surface field (BSF) in monofacial design. The configuration for the rear metallization and contact is illustrated on Figure 2. To accurately establish optical parameters, the wavelength-dependent external front surface transmission ($T_{ext}$) was determined using OPAL2 [6]. The front side layers consisted of an n+ type poly-Si layer and a thin silicon oxide layer, with thicknesses of 70 nm and 1.5 nm, respectively. The choice of the TOPCon stack on the front side is guided by the need to balance its optical and passivation properties. A key consideration is that the poly-Si layer can degrade during the high-temperature processes typically required for rear-side metal contact formation. To mitigate this, a successful solution has been developed by introducing a sacrificial layer that prevents degradation [7].

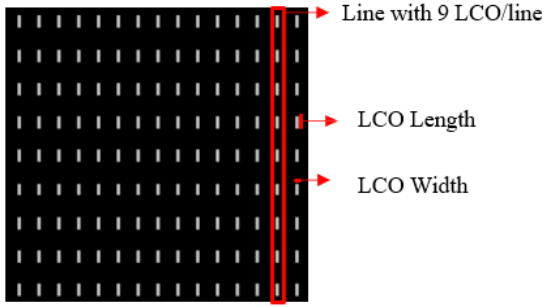

**Figure 2:** Rear Side Metal and Contact Configuration

The optical model used in this study is based on the standard AM 1.5G spectrum. Initially, the current-voltage (*I-V*) characteristics of the structure were calculated under 1-Sun irradiance. However, this does not represent practical conditions, as the bottom cell in a perovskite/silicon tandem configuration does not absorb the full spectrum [8]. To address this, a second set of simulations was conducted under 0.5-Sun irradiance. In both scenarios, the main goal is to assess the impact of the rear-side metal on $V_{oc}$ and FF values. The study aims to identify the optimal metal fraction (f) based on these parameters, without considering the effects on short-circuit current and power conversion efficiency.

To optimize the rear side contact, a two-step process was employed as detailed in Table II. This approach allowed for a comprehensive exploration of the impact of key parameters on the metal fraction and facilitated the identification of optimal conditions for the hybrid PERC/TOPCon structure.

**Table II:** Optimization Plan

|  | Parameters |
|---|---|
| **STEP I** f = 0.6-4.1% | LCO/line = 24 LCO Length = 400 μm Number of lines = 10-35 LCO Width = 30-60 μm |
| **STEP II** f = 1.4-3% | Optimal values in STEP I are kept constant LCO/line = 17-31 LCO Length = 350-410 μm |

## 3 RESULTS AND DISCUSSION

To comprehensively understand the impact of rear-side metal configuration on the performance parameters $V_{oc}$ and FF, Figure 3 and Figure 4 illustrate Step I and Step II under 1 Sun illumination, while Figure 5 and Figure 6 depict the same steps under 0.5 Sun illumination. To explore the trade-off between $V_{oc}$ and FF as metal fraction increases, their product was calculated as well.

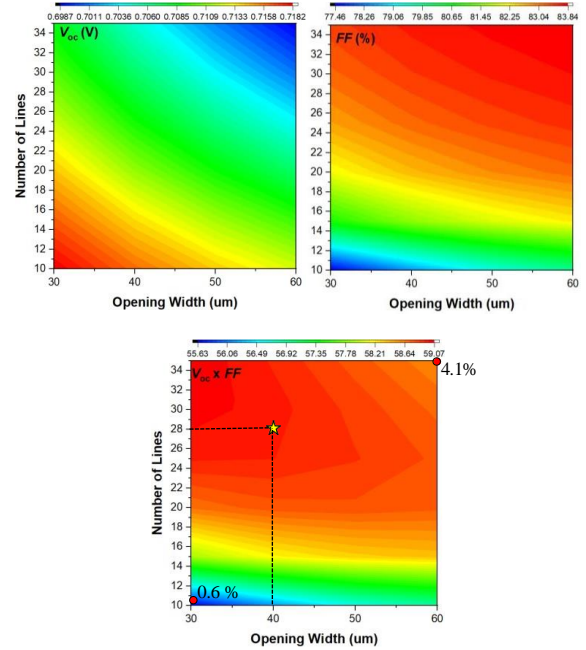

**Figure 3:** $V_{oc}$, FF and $V_{oc}$ x FF dependency on opening width and number of lines for a fixed number of LCO/line of 24 with fixed length of 400 μm (Step I) under 1 Suns illumination

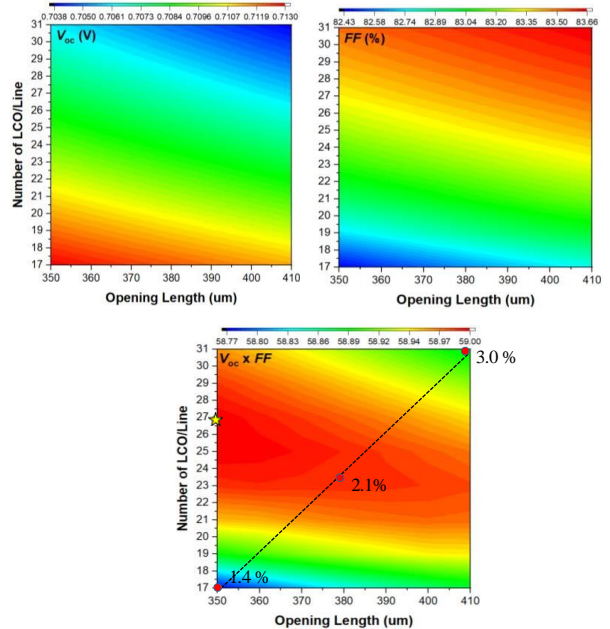

**Figure 4:** $V_{oc}$, FF and $V_{oc}$ x FF dependency on number of LCO/line and opening length for fixed number of lines of 28 and opening width of 40 um (Step II) under 1 Suns illumination

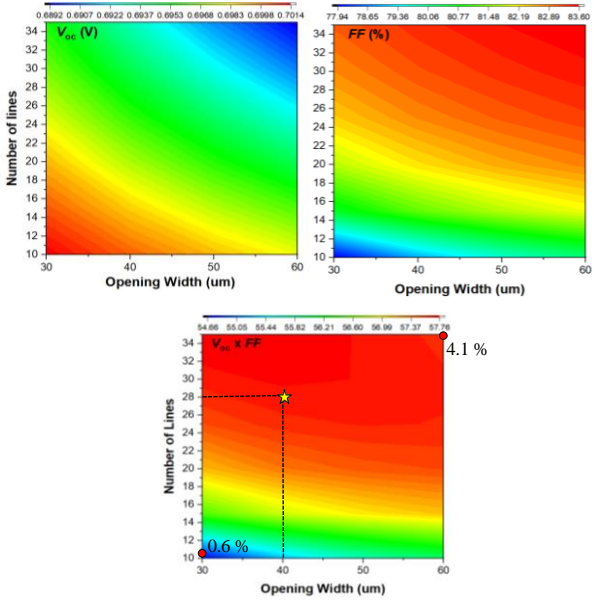

**Figure 5:** $V_{oc}$, FF and $V_{oc}$ x FF dependency on opening width and number of lines for a fixed number of LCO/line of 24 with fixed length of 400 μm (Step I) under 0.5 Suns illumination

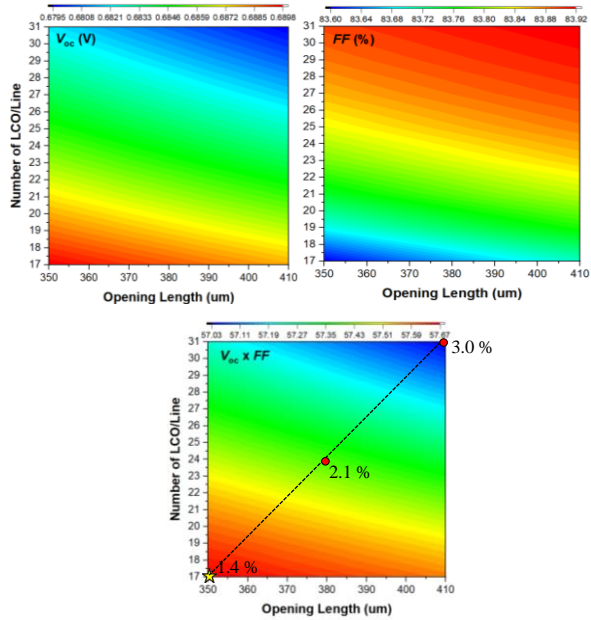

**Figure 6:** $V_{oc}$, FF and $V_{oc}$ x FF dependency on number of LCO/line and opening length for fixed number of lines of 28 and opening width of 40 um (Step II) under 0.5 Suns illumination

Under 1 Sun illumination, the behavior of both $V_{oc}$ and FF shows a clear pattern as the metal fraction increases. Specifically, $V_{oc}$ experiences a steady decline, while FF increases, reflecting a typical trade-off between these two performance parameters. However, when the two steps are analyzed separately, it becomes evident that the product of $V_{oc}$ × FF does not exhibit the same trend across both steps in relation to the metal fraction. In Step I, FF reaches a saturation point at approximately f ≈ 1%, indicating that further increases in the metal fraction beyond this value yield diminishing improvements in FF. On the other hand, in Step II, the saturation point occurs later, at around f ≈ 2.5%, suggesting that FF can be further optimized with a higher metal fraction in this step.

The indications for these behaviors are shown in the production plots as well, where the product of both parameters is fully dominated by the FF in the first step. As the opening length and their pitch on the second step change, a higher trade-off between the series resistance loss and rear shading area which corresponds to the losses in $V_{oc}$ and gains in FF are seen [6]. This trade-off is balanced with a metal fraction of 2.02%. This metal fraction represents the optimal choice between the $V_{oc}$ losses and FF gain. However, the current literature indicates that the optimal metal fraction for conventional PERC typically falls within the 3-5% range. In this case, a lower optimal metal fraction is observed, likely due to the front emitter's excellent surface passivation (low $J_0$) compared to the diffused emitter of PERC.

For the 0.5 Sun case, Step I mirrors the trends seen under 1 Sun, with similar patterns of $V_{oc}$ reduction and FF increase as the metal fraction increases. However, Step II exhibits a distinct shift in behavior compared to the 1 Sun case. Unlike the 1 Sun case where FF dominates, under 0.5 Sun illumination, $V_{oc}$ becomes the key determining factor in the trade-off. This indicates that at lower illumination levels, the effects of series resistance and other electrical losses become more pronounced, making $V_{oc}$ the critical parameter to optimize in Step II under reduced light conditions. Despite these differences, the analysis under both illumination levels reinforces the importance of fine-tuning the metal fraction to achieve the best balance between $V_{oc}$ and FF across both optimization steps.

## 4 CONCLUSIONS

To conclude, the presented study underscores the significance of optimizing the rear side metal configuration in the TOPCon/PERC hybrid structure. The insights gained from this two-step process shed light on the balance $V_{oc}$ and FF, emphasizing the importance of parameters such as opening width, number of lines, the number of openings per line and length of local opening contacts for achieving optimal photovoltaic performance.

ACKNOWLEDGEMENT: Funded by the European Union. Views and opinions expressed are however those of the author(s) only and do not necessarily reflect those of the European Union or RIA. Neither the European Union nor the granting authority can be held responsible for them. NEXUS project has received funding from the European Union's Horizon Europe research and innovation program under grant agreement No. 101075330.